\pdfoutput=1

\documentclass[a4paper,11pt]{article}

\usepackage{jcappub} 
\usepackage{color}
\usepackage{xcolor}
\usepackage{amsmath}
\usepackage{amsfonts}
\usepackage{graphicx} 
\usepackage{amssymb}
\usepackage{epstopdf}
\usepackage{url}
\usepackage{bm}
\usepackage{hhline}
\usepackage{multirow}
\usepackage{epsfig}
\usepackage{varioref}
\usepackage{amssymb}
\usepackage{comment} 
\usepackage{cases}

\usepackage{multirow}
\usepackage{mathrsfs}
\usepackage{calrsfs}
\usepackage{mathtools}
\usepackage{amsthm}
\usepackage{hyperref}
\usepackage{wrapfig}
\usepackage{caption}
\usepackage{subcaption}

\include{configs} 

\title{{\it Planck} Satellite Constraints on Pseudo-Nambu--Goldstone Boson Quintessence}

\author{Vanessa Smer-Barreto and Andrew R. Liddle} 

\affiliation{Institute for Astronomy, University of Edinburgh, Royal Observatory, Blackford Hill, Edinburgh EH9~3HJ, United Kingdom}

\emailAdd{vsm@roe.ac.uk, arl@roe.ac.uk}

\date{\today}

\abstract{
The Pseudo-Nambu-Goldstone Boson (PNGB) potential, defined through the amplitude $M^4$ and width $f$ of its characteristic potential $V(\phi) = M^4[1 + \cos(\phi/f)]$, is one of the best-suited models for the study of thawing quintessence. We analyse its present observational constraints by direct numerical solution of the scalar field equation of motion. Observational bounds are obtained using Supernovae data, cosmic microwave background temperature, polarization and lensing data from {\it Planck}, direct Hubble constant constraints, and baryon acoustic oscillations data. We find the parameter ranges for which PNGB quintessence gives a viable theory for dark energy. This exact approach is contrasted with the use of an approximate equation-of-state parametrization for thawing theories. We also discuss other possible parameterization choices, as well as commenting on the accuracy of the constraints imposed by {\it Planck} alone. Overall our analysis highlights a significant prior dependence to the outcome coming from the choice of modelling methodology, which current data are not sufficient to override. 
}
\begin{document}

\maketitle
\section{\label{PNGB:intro} Introduction }
The nature of the Universe's late-time acceleration, originally discovered via measurements of 51 type Ia Supernovae  \cite{Riess, Perlmutter}, remains unexplained. 
Under Einstein's general relativity scheme,  astronomical observations \cite{PlanckCos15,PlanckDE15} indicate that 68\% of the energy density of the Universe corresponds to dark energy, a material component with negative pressure and the originator of the poorly-understood accelerated cosmological expansion at present. Reference~\cite{Copeland:2006wr} is a comprehensive review on the subject, providing a good number of candidate dark energy models.

A plausible choice for explaining the characteristics of dark energy is the cosmological constant $\Lambda$, appealing for its simple  equation of state $w = -1$ and its accurate agreement with observations. However, this choice has awkward implications, in particular the $120$ orders of magnitude difference between the expected theoretical value of the vacuum energy density and the observational magnitude of $\Lambda$, which  raises concern. The so-called cosmic coincidence problem, in reference to the circumstance that both the matter and dark energy components of the Universe are of the same order of magnitude at the current epoch, despite having been so different at every other stage of the cosmos's history, adds more doubts to the underpinning of the standard cosmological model. The increasing amount of available data and the aforementioned theoretical dilemmas have promoted the discussion of alternative possibilities to the cosmological constant in order to deepen our understanding of the present and future state and composition of the Universe. One of those possibilities is an evolving dark energy equation of state.  
 
A canonical scalar field which varies slowly along a potential $V(\phi)$, known as quint\-essence, can lead to the very similar observational results as the cosmological constant, with the advantage of a broader phenomenology and the possibility of a link to fundamental physics models. Reference~\cite{Tsuji} reviews the cosmological dynamics of quintessence, providing approximate analytical solutions wherever possible. These theories can be approximately divided into two classes: tracking type, in which the equation of state decreases towards $w = -1$ as accelerated expansion commences, and thawing type, in which the field is initially frozen by Hubble friction during radiation and matter domination until late times when it is finally allowed to roll down the potential. 
 
In this paper we analyze one of the best candidates for thawing quintessence, the Pseudo Nambu--Goldstone Boson (PNGB) model \cite{Stebbins}. The PNGB's characteristic potential is sufficiently flat as to give an appropriately-negative equation of state with $w$ close to $-1$. Previous efforts to constrain PNGB quintessence may be found in Refs.~\cite{Dutta,Waga,Ng,Kawasaki}, while Ref.~\cite{Abrahamse} carried out forecasts on the ability of future experiments to constrain this model. An alternative approach taken into the study of thawing quintessence is via a suitable approximate equation of state, established in Refs.~\cite{DuttaAndScherrer,ChibaFirst}, where the parametrization is done via the present equation of state $w_0$, the field density parameter, and the gradient of the selected potential. A parameter estimation analysis using this method has previously been carried out in Ref.~\cite{Chiba}. The PNGB potential constraints have also been thoroughly analysed in the context of dark matter densities in Ref.~\cite{Hlozek:2014lca}.

This paper is organized as follows. In Sec.~\ref{PNGB:background} we present the background evolution equations of PNGB quintessence. In Sec.~\ref{PNGB:cosmomc} we describe our analysis by direct parametrization of the width and amplitude of the potential, performed with CosmoMC version July 2015 and the quintessence module included within the CAMB code \cite{CAMB}. We mention the prior range for the model parameters and the datasets utilised. Section~\ref{PNGB:results} contains the observational bounds obtained for this model. The accuracy of the approximate approach developed in Refs.~\cite{DuttaAndScherrer,ChibaFirst} is analysed in Sec.~\ref{PNGB:eqstate} with our full data combination. The performance of {\it Planck} data against the {\it Planck}, JLA, HST and BAO combination is discussed in Sec.~\ref{PNGB:discussion:Planck}, followed by a discussion of the choice of priors in this particular model in Sec.~\ref{PNGB:discussion:priors} and a comparison against previous studies of the same potential in Sec.~\ref{PNGB:discussion:previous}. We finish in Sec.~\ref{PNGB:conclusions} with some concluding remarks on this work.
\section{\label{PNGB:background} PNGB quintessence}
We assume a spatially-homogeneous quintessence field described by the scalar $\phi$ and its potential $V(\phi)$. The action is
\begin{equation}
S = \int d^4x \sqrt{-g} \left[ \frac{1}{2}M_{\mathrm{Pl}}^2 R - \frac{1}{2}g^{\mu \nu} \partial_{\mu} \phi \partial_{\nu} \phi - V(\phi)\right] + S_\mathrm{m} \,.
\label{eq:Action}
\end{equation}
where $g$ is the determinant of the metric $g_{\mu \nu}$, $R$ is the Ricci scalar, $M_{\mathrm{Pl}}$ is the reduced Planck mass, and $S_\mathrm{m}$ is the action for non-relativistic matter.

For a statistically spatially homogeneous and isotropic universe with Friedmann--Robert\-son--Walker (FRW) metric, $ds^2 = -dt^2 + a^2(t) d\vec{x}^2$, the background evolution is given by
\begin{eqnarray}
\ddot{\phi} + 3H\dot{\phi} + \frac{dV}{d\phi}  & = & 0\,.
\label{eq:Field} \\
3H^2 M_{\mathrm{Pl}}^2 & =& \rho_{\phi} + \rho_\mathrm{m}\,.
\label{eq:Friedmann}\\
\dot{\rho}_\mathrm{m} + 3H\rho_\mathrm{m} & = & 0 \,.
\label{eq:Fluid}
\end{eqnarray}
where $H = \dot{a}/a$ is the Hubble parameter and a dot stands for a derivative with respect to physical time $t$. The pressure and energy density of the scalar field are given by $P_{\phi} = \dot{\phi}^2/2 - V(\phi)$ and $\rho_{\phi} = \dot{\phi}^2/2 + V(\phi)$ respectively. The quintessence field equation of state is defined as $w \equiv P_{\phi}/\rho_{\phi}$. The density parameter is $\Omega_\phi = \rho_{\phi}/\rho_{\rm crit}$, where $\rho_{\rm crit}$ is the critical density of the Universe. 
\begin{figure}
        \centering
                \includegraphics[width=0.7 \textwidth]{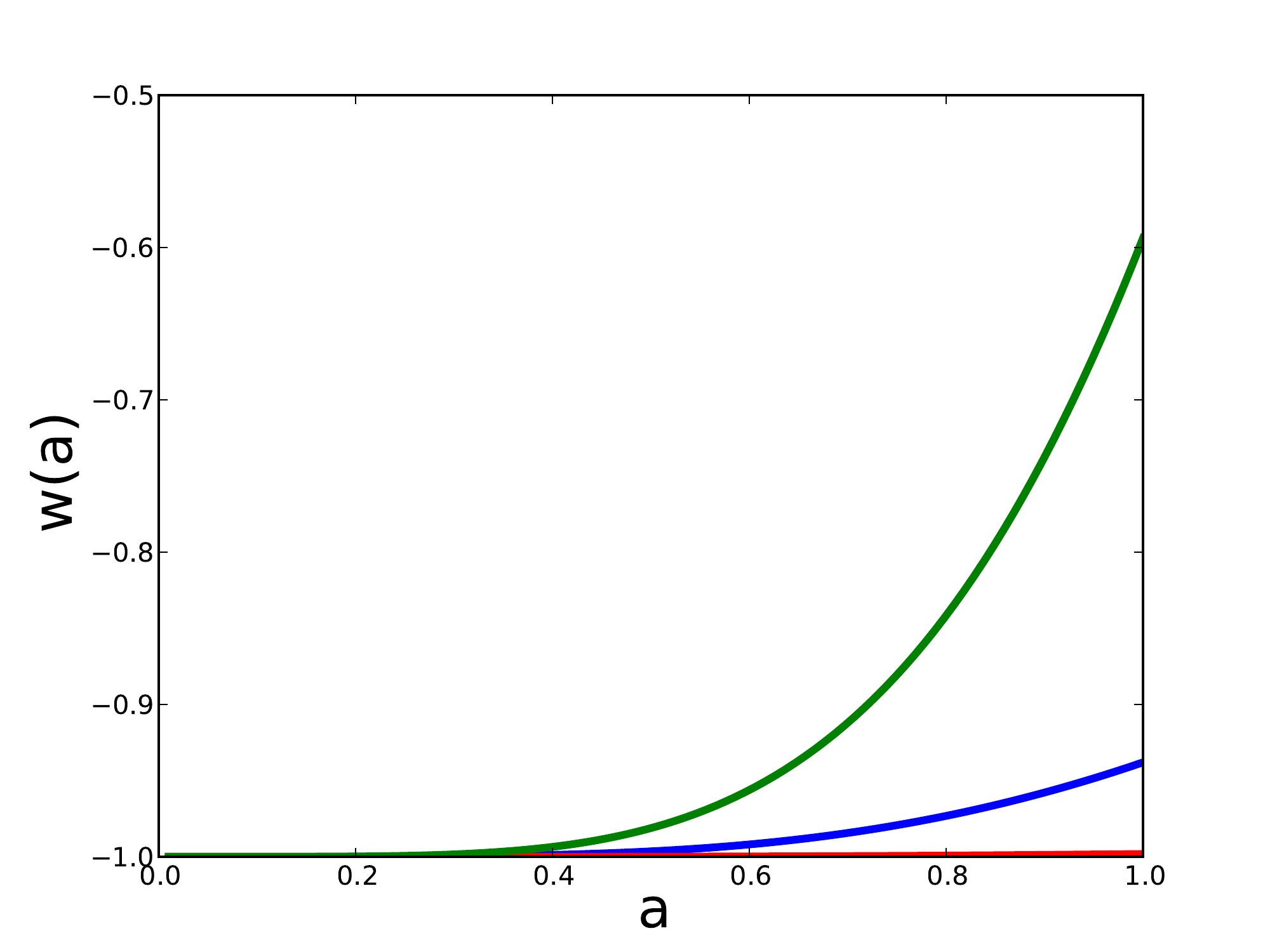}
                \label{fig:EqState1}
        \caption{Equation of state versus scale factor for the PNGB potential with $f = 1.4 M_{\rm Pl}$, for models leading to present values $w_{\rm 0} = -0.99$ (red), $w_{\rm 0} = -0.93$ (blue) and $w_{\rm 0} = -0.6$ (green). In thawing quintessence, $w \geq -1$.}\label{fig:Potential}
\end{figure}
 The model under consideration is of the thawing type; which means it has a potential able to mimic a nearly-frozen field during the matter-dominated era, caused by Hubble friction, implying $w$ near $-1$ at early times. Its kinetic energy contribution must be kept small, which means a small mass is required as well as a nearly-flat potential. The PNGB theory fits all of these conditions. Its potential is
 \begin{equation}
V(\phi) = M^4 \left[1 + \cos  \frac{\phi}{f} \right] \,,
\label{potential}
\end{equation}
where $M^4$ is the amplitude of the potential and $f$ determines the width of the function. Figure~\ref{fig:Potential} shows some sample evolutions of the equation of state obtained numerically for models giving present density parameter $\Omega_\phi = 0.68$. The dynamics and motivational background of the model are analyzed in detail in Ref.~\cite{Stebbins}.
 
The parameters that determine the cosmological evolution of this model are its normalization $M^4$, its width $f$, and the initial conditions $\phi_\mathrm{i}$ and $\dot{\phi_\mathrm{i}}$.  The rapid early expansion of the Universe leads to sharply-decaying field velocity at early times, enabling us to assign $\dot{\phi} = 0$ as the initial condition for our numerical evolution. Requiring that at present the quintessence field has a particular density parameter $\Omega_{\phi}$ allows us to use the density parameter as a variable and treat $\phi_\mathrm{i}$ as a derived parameter. The background evolution is calculated from a scale factor $a = 10^{-9}$. 
\section{\label{PNGB:cosmomc} PNGB potential analysis with CosmoMC}
\begin{table}[t]
\centering
\begin{tabular}{|c|c|}
 \hline
 Parameter & Prior range  \\ \hline
 $f/M_{\rm Pl}$ & [0.1, 2]   \\
 $M^4$ & [0.25, 2]  \\
 $\Omega_{\rm b} h^2$ & [0.005, 0.1] \\
 $\Omega_{\rm c} h^2$ & [0.001, 0.99] \\
 $n_{\rm s}$ & [0.8,1.2] \\
 $\tau$ & [0.01, 0.8] \\
 $100 \theta_{\rm MC}$ & [0.5, 10] \\
 $\ln(10^{10}A_{\rm {s}})$ & [2, 4] \\
 $w_0$ & ...\\
 $\Omega_{\phi}$ & ...\\
 $\phi_{\rm i}/f$ & ...\\
 \hline
\end{tabular}
\caption{\label{AllPars} Prior ranges for cosmological and PNGB model parameters, the prior being uniform in the parameter quoted. The meaning of the cosmological parameters is as in the {\it Planck} collaboration papers \cite{PlanckCos15,PlanckLikelihood}. The final three parameters listed are derived from the others and inherit non-uniform priors from their relation to them.}
\end{table}
The exploration of the parameter space for the PNGB potential was made using the CosmoMC program, version July 2015 \cite{CosmoMC}. Within the CAMB program the quintessence module was updated by us to be compatible with the CAMB/CosmoMC version of July 2015. Perturbations in the quintessence field are fully taken into account for the calculation of the cosmological observables. Additionally, a modification of the original code was implemented to solve a starting point issue: in order for the evolution to commence, the amplitude of the potential has to be large enough to allow the quintessence density parameter $\Omega_{\phi}$ to have a value corresponding to the observed dark energy density. The random search nature of the MCMC code (at least at the beginning of the parameter space exploration) caused the program to stop through failing to meet this condition, therefore not allowing the code to calculate a likelihood. Instead of stopping the code after an unsuccessful initial setting, the unsuitable parameters were assigned an improbable negative logarithmic likelihood of $10^{30}$ (the standard value for the program to deem a set of parameters unlikely) therefore rejecting them but allowing the rest of the estimations to continue.

The choice of prior ranges for the standard cosmological parameters were taken as in the {\it Planck} collaboration 2015 analysis \cite{PlanckCos15}. These, as well as the added free components that characterize the PNGB model, are displayed in Table~\ref{AllPars}. Additionally we impose a range on $H_{\rm{0}}$ of $50\, {\rm km}\,{\rm s}^{-1}\,{\rm Mpc}^{-1} \leq H_0 \leq 80\,{\rm km}\,{\rm s}^{-1}\,{\rm Mpc}^{-1}$. 

Previous related works \cite{Dutta,Abrahamse} have imposed a hard limit on the width of the potential of $f < M_\mathrm{Pl}$, citing both computational reasons (avoiding a divergent direction for the MCMC chains to reach convergence) and theoretical ones, such as inaccuracy of the described potential for $f > M_\mathrm{Pl}$ and motivations from string theory. Referring to the latter, the PNGB potential has been used with the purpose of understanding natural inflation better. One of the main points of this approach has focused on the different values of the potential's width $f/M_{\mathrm{Pl}}$, in the context of supersymmetry/superstring theories. The overall conclusion favours $f < M_{\mathrm{Pl}}$ \cite{Banks, Abrahamse}. However we choose to impose a somewhat weaker prior $f < 2 M_{\rm Pl}$, by which time the potential is flat enough that it can commonly generate observables practically indistinguishable from those for $\Lambda$CDM. Our motivation to do this is because supersymmetric theories have recently become less compelling, given the lack of evidence in their direction, and moreover the PNGB potential offers a phenomenological description of dark energy that is well behaved, independently of the use of axions during the inflationary epoch. We will study the outcome of choosing alternative priors for $f$ later.

To obtain the observational bounds we utilized the JLA compilation of supernova distances \cite{JLA}, cosmic microwave background temperature anisotropies, lensing and polarization data from the {\it Planck} 2015 data release \cite{PlanckCos15, PlanckLikelihood}, direct constraints on the Hubble constant from Hubble Spacee Telescope observations \cite{HST}, and baryon acoustic oscillations data from SDSS~\cite{BAO11}.
\section{\label{PNGB:results} Results}
\begin{figure}
\centering
\begin{subfigure}{.5\textwidth}
  \centering
  \includegraphics[width=1\linewidth]{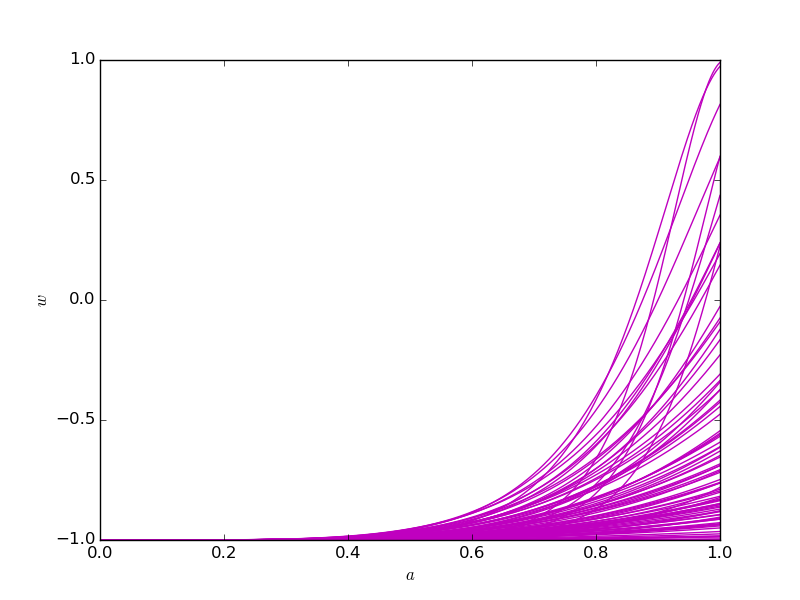}
\end{subfigure}%
\begin{subfigure}{.5\textwidth}
  \centering
  \includegraphics[width=1\linewidth]{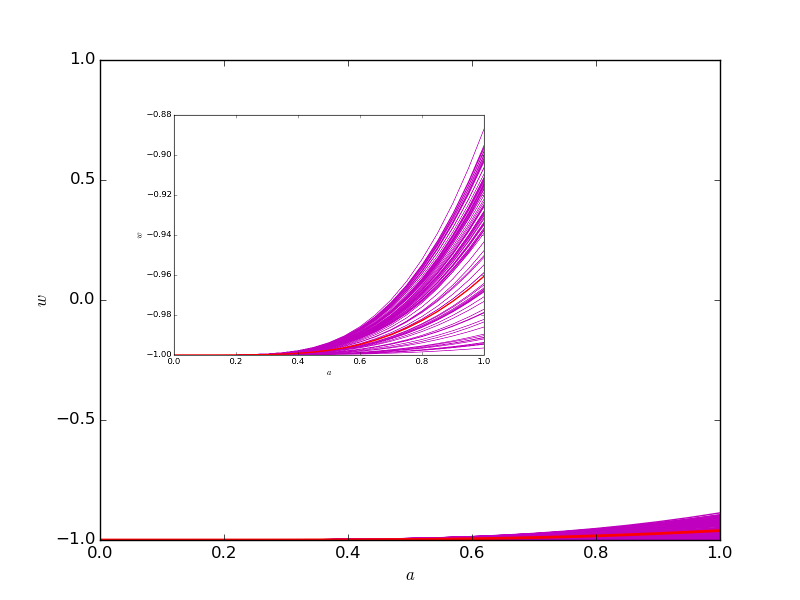}
\end{subfigure}
\caption{Density plots for the equation of state $w$ versus scale factor $a$ of the PNGB model. The figure on the left depicts random prior choices from Table~\ref{AllPars}. On the right, a sample of the posterior distribution models obtained from the combined JLA + BAO + HST + \textit{Planck} 2015 datasets is shown. The best-fit model is drawn in red, and a zoom is included to show the detail of the posterior trajectories given their closeness on the original axis range.}
\label{fig:densityplots}
\end{figure}

The free parameters of our analysis, aside from those standard in any cosmological model, are the width $f$ and the amplitude $M^4$ of the potential (\ref{potential}). The CAMB code uses dimensionless versions of the field and width parameter, but we will refer to them in their native units of mass. The code's background evolution computation handles all energy--momentum components with a normalization equivalent to $8\pi G \rho_{\mathrm{crit}} /3 c^2$ where $\rho_{\mathrm{crit}}$ is the present value of the critical density. The matter energy constituents are then multiplied using the standard density parameter definition $\Omega_{\rm i} = \rho_\mathrm{i} / \rho_{\mathrm{crit}}$ where the index $\rm {i}$ is matter or quintessence. This means that $M^4$ and $V(\phi)$ are, regarding their units, equivalent to the density parameter for the quintessence field $\Omega_{\phi}$. 

Figure~\ref{fig:densityplots} depicts an ensemble of trajectories for the equation of state $w$ versus scale factor $a$ of the PNGB model, drawn from both the prior and posterior distributions. We see a very strong tightening of the posterior distribution with respect to the prior, indicating that the data are significantly constraining the considered set of models. 

\begin{figure}[t]
        \centering
                \includegraphics[width=.99 \textwidth]{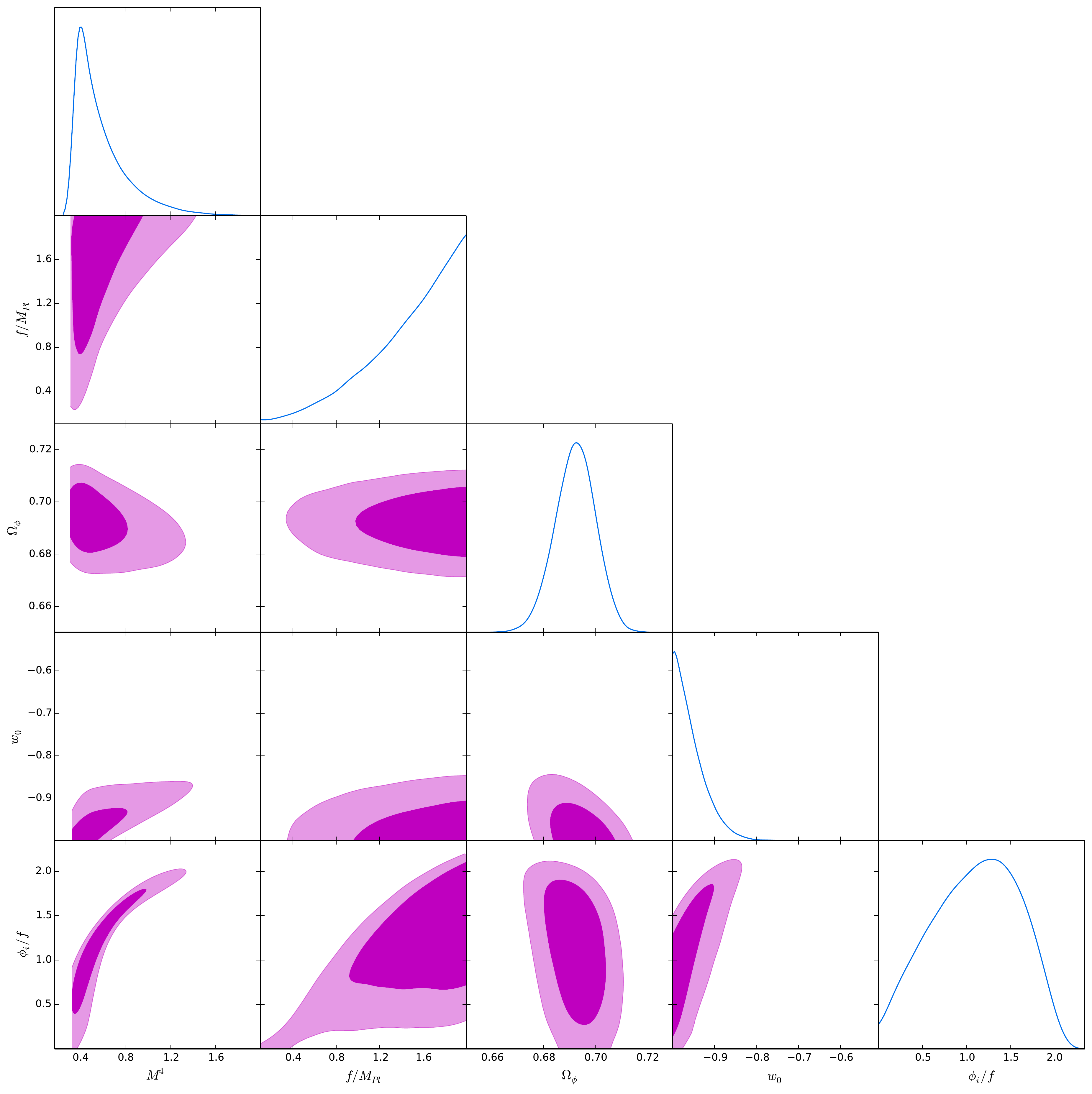}
       \caption{2D contours of the combined JLA + BAO + HST + {\it Planck} 2015 constraints for the PNGB model with potential Eq.~(\ref{potential}). The  individual marginalized posterior probability distributions of each parameter are also shown.} \label{fig:qStandardAll}
\end{figure}

In Fig.~\ref{fig:qStandardAll}, a triangular plot with the  68\% and 95\% confidence contours between the amplitude $M^4$, the width $f/M_{\rm{Pl}}$, the field density parameter $\Omega_{\phi}$, the present value of the field equation of state $w_0$ and the initial displacement of the field $\phi_{\rm i}/f$ is displayed. The individual posterior distributions of each parameter are also shown.

The parameter  $f/M_{\rm Pl}$ is unconstrained at the upper edge of its prior, and has 68\% and 95\% confidence lower limits of $1.34$ and $0.67$ respectively. For $M^4$ the 68\% and 95\% confidence ranges are $M^4 = 0.58_{-0.26}^{+0.07}$ and $M^4 = 0.58_{-0.31}^{+0.52}$. These probability distributions are expected. As $f$ grows the potential flattens, returning the cosmological constant case, which makes it impossible to confine this parameter at its upper value, while $M^4$ is sharply cut off at the lowest value able to sustain an allowable present density parameter, while fitting the data well at that value. Note however that the quoted lower limits on $f$ depend strongly on the assumed {\it upper} limit of its prior, as models beyond the adopted prior continue to fit the data well. We make a detailed analysis of prior dependence in Sec.~\ref{PNGB:discussion}.

The density parameter $\Omega_{\phi}$ is constrained at 68\% and 95\% confidence as, $\Omega_{\phi} = 0.69 \pm 0.01$ and $\Omega_{\phi} = 0.69_{-0.02}^{+0.01}$ respectively. The 95\% upper limit on the present equation of state is $w_0 = -0.88$, whilst the lower cut-off of $w_0 = -1$ reaffirms the cosmological constant limit of this theory. This range is as expected for thawing quintessence; while we have permitted values of the present equation of state bigger than $-1$, allowing non-trivial past dynamics for dark energy, the cosmological constant case gives a very good fit to the astronomical data.

Examination of $\phi_\mathrm{i}/f$ gives a criterion for the slope of the potential at the start of the cosmological evolution. For $\phi_\mathrm{i}/f \approx 0$ the feature at the bottom left corner of the ($f/M_{\rm{Pl}}$, $\phi_\mathrm{i} / f$) duplet shows that only narrow potentials (those corresponding to small $f$) are allowed. In such a steep regime, a slope slightly different from zero would cause the quintessence field to evolve too quickly, therefore not reproducing a thawing behaviour. Once the slope increases to $\phi_\mathrm{i} /f \approx 0.3$ most of the $f$ range is enabled. Larger values for $\phi_\mathrm{i} / f$ are favoured when $f$ is larger. We find a 95\% upper limit of $\phi_\mathrm{i} /f = 1.9$, while there is no lower limit since $\phi_\mathrm{i} /f = 0$ reproduces $\Lambda$CDM precisely. The upper boundary value shows there is a fair range of models permitted with an evolving scalar field. However this regime is typically not well recovered except for narrow potentials with $f \approx 0$, because in order to attain such it is also necessary for $M^4$ to be very close to $\Omega_{\rm v}$, meaning only a narrow sliver of prior space is available. The data clearly favours the direction of increasing $f$. 
\section{ \label{PNGB:eqstate} Thawing quintessence: equation of state parameterization}
An alternative way of parametrizing thawing quintessence behaviour is by use of an approximate analytical expression for its equation of state. Such a study has been carried out in this manner in Ref.~\cite{Chiba}, our main new contribution being the use of more recent data (particularly {\it Planck} 2015) and a comparison with the exact evolution of the PNGB potential. The derivation of this solution can be found in Refs.~\cite{DuttaAndScherrer, ChibaFirst, Tsuji}, and leads to the equation of state
\begin{equation}
w(a) = -1 + \left(1 + w_0\right)a^{3(K-1)} \left[ \frac{(K - F(a))(F(a) +1)^K + (K+F(a))(F(a)-1)^K}{(K-\Omega_{\rm \phi}^{-1/2})
(\Omega_{\rm \phi}^{-1/2}+1)^K + (K + \Omega_{\rm \phi}^{-1/2})(\Omega_{\rm \phi}^{-1/2} - 1)^K}   \right]^2,
\label{ThawingEqState}
\end{equation}
where 
\begin{eqnarray}
K & = & \sqrt{1 - \frac{4M_{\rm Pl}^2 V_{\rm ,\phi \phi}(\phi_\mathrm{i})}{3V(\phi_{\rm i})}},
\label{K}\\
F(a) & = & \sqrt{1+ (\Omega_{\rm \phi}^{-1} - 1)a^{-3}}.
\label{Feq} 
\end{eqnarray}
\begin{figure}[t]
        \centering
                \includegraphics[width=0.7 \textwidth]{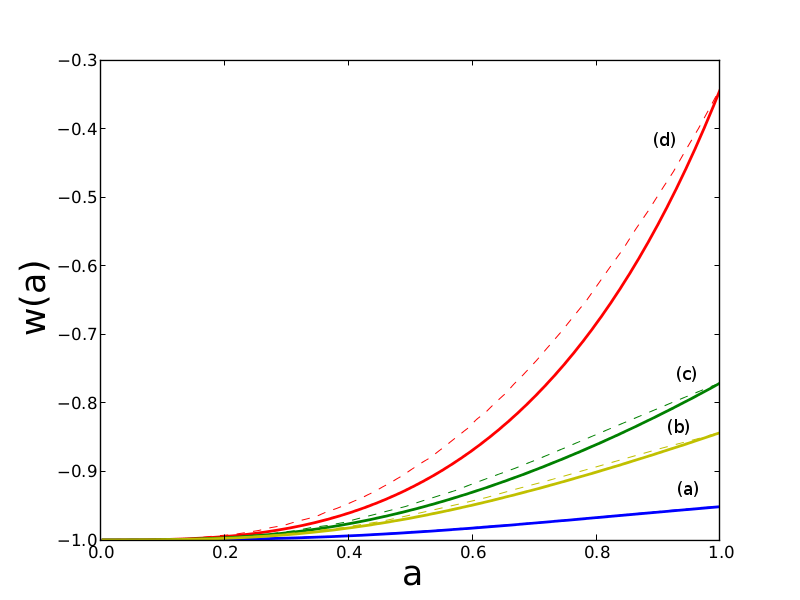}
 \caption{ Equation of state versus scale factor for the quintessence field equation of state (solid lines) and Eq. (\ref{ThawingEqState}) (dashed lines) with $\Omega_{\rm \phi} = 0.68$. The parameter values in each case are (a) $f/M_{\rm{Pl}} = 2$, $\phi_{\rm{i}}/f = 1.65$, $K = 0.98$, $w_{\rm{0}} = -0.95$; (b) $f/M_{\rm{Pl}} = 1.1$, $\phi_{\rm{i}}/f = 1.52$, $K = 1.02$, $w_{\rm{0}} = -0.84$; (c) $f/M_{\rm{Pl}} = 0.9$, $\phi_{\rm{i}}/f = 1.43$, $K = 1.09$, $w_{\rm{0}} = -0.77$; (d) $f/M_{\rm{Pl}} = 0.5$, $\phi_{\rm{i}}/f = 0.94$, $K = 1.72$, $w_{\rm{0}} = -0.34$.}                \label{fig:comovingWs}
\end{figure}
The equation of state (\ref{ThawingEqState}) is expressed in terms of three parameters: $\Omega_{\rm \phi}$ at present, $w_{\rm 0}$, and $K$ which measures the curvature of the scalar field potential at its maximum \cite{Tsuji}. For $K$ larger than 10 the movement of the field at the start of the evolution is required to be very small to avoid a quick roll down. If the field touches the minimum of the potential and starts oscillating at a scale factor value near today's, numerical simulations establish that Eq.~(\ref{ThawingEqState}) is not valid anymore. In addition to this inaccuracy, for an oscillating potential the equation of state would become positive, therefore violating the $w < -1/3$ condition for a dark energy description. For $K$ smaller than 0.5 the field mass becomes very large, implying that the Taylor expansion around $\phi = \phi_{\rm i}$ becomes inaccurate because of the rapid variation of the field. We have a particular focus on finding a confidence range for the curvature of the potential $K$ that is better constrained than in past analyses by making use of the exact results in Sec.~\ref{PNGB:results}.
 
In Fig.~\ref{fig:comovingWs} we compare the behaviour of the thawing equation of state against the numerical solution provided by the quintessence module of CAMB. The approximation works very well for $w_0 \approx -1$ and $K \approx 1$, but becomes less accurate for larger values of $w_0$ which in turn correspond to smaller values of $K$. If instead we had adjusted curves generated by Eq.~(\ref{ThawingEqState}) to the best-fit values of $K$ and $w_{\rm{0}}$ we would have possibly found a more varied set of parameters able to duplicate a broader set of solutions. 

\begin{figure}[t]
        \centering
                \includegraphics[width=0.45 \textwidth]{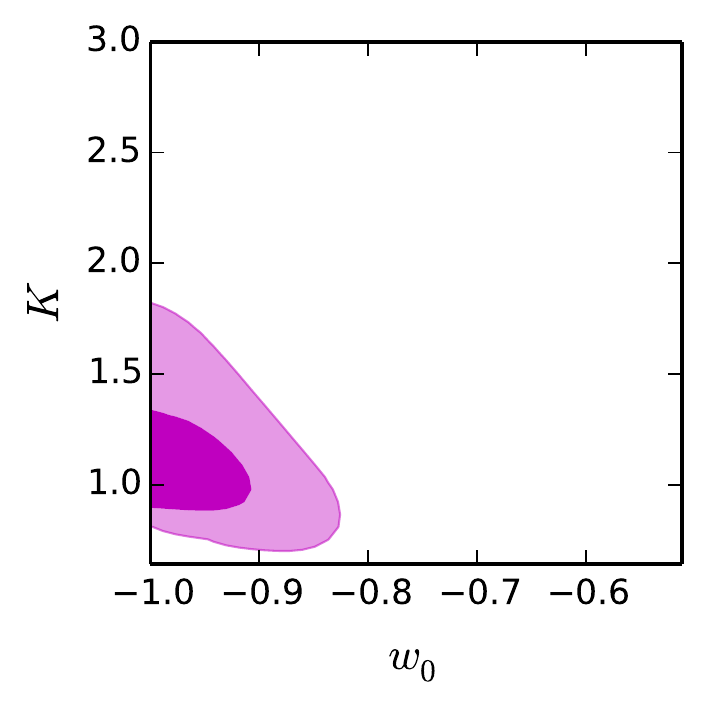}
        \caption{The 68\% and 95\% confidence level regions for ($w_0, K$) given by Eq.~(\ref{K}). } \label{fig: K_w_2D}
\end{figure}

In Fig.~\ref{fig: K_w_2D} the confidence contours for the curvature parameter $K$ are shown. The 95\% constraint obtained is $K = 1.1 \pm 0.4$, where $K$ was calculated as a derived parameter from the results of Sec.~\ref{PNGB:results}. Given the condition within the CAMB module Quint to attain convergence of $\Omega_{\phi}$ before allowing the evolution of $\phi_i$ to commence, the width of the potential dictates the value of the curvature. As we discuss in Sec.~\ref{PNGB:discussion:previous}, the distribution of $K$ values that emerges from simulating the PNGB potential is far from uniform, and hence our constraints appear very different from those that sample uniformly in $w_0$ and $K$ such as Ref.~\cite{Chiba}. In Ref.~\cite{Chiba}, the authors make a distinction between values of $K<1$, where they obtain a 95\% estimate of $-1.21 < w_0 < -1.00$, and $K<2$ which yields $-1.14 < w_0 < -0.92$. These values fall largely into the phantom region $w < -1$, and are only partially compatible with ours. However, in their case, $K$ is allowed to grow up to a value of $9.95$, giving a $-2.06 < w_0 < -1.01$. From here we can conclude that constraining the cosmological trajectories with a specific potential makes a big difference in preventing this type of models from suffering instabilities. 
\section{ \label{PNGB:discussion} Discussion}
In this section we place the results of Secs.~\ref{PNGB:results} and~\ref{PNGB:eqstate} in context by comparing them to results under different choices of datasets and parameter priors, and then compare to constraints obtained by previous authors on thawing quintessence.
\subsection{ \label{PNGB:discussion:Planck} Constraints using {\it Planck} 2015 only}
\begin{figure}[t]
        \centering
                \includegraphics[width=0.99 \textwidth]{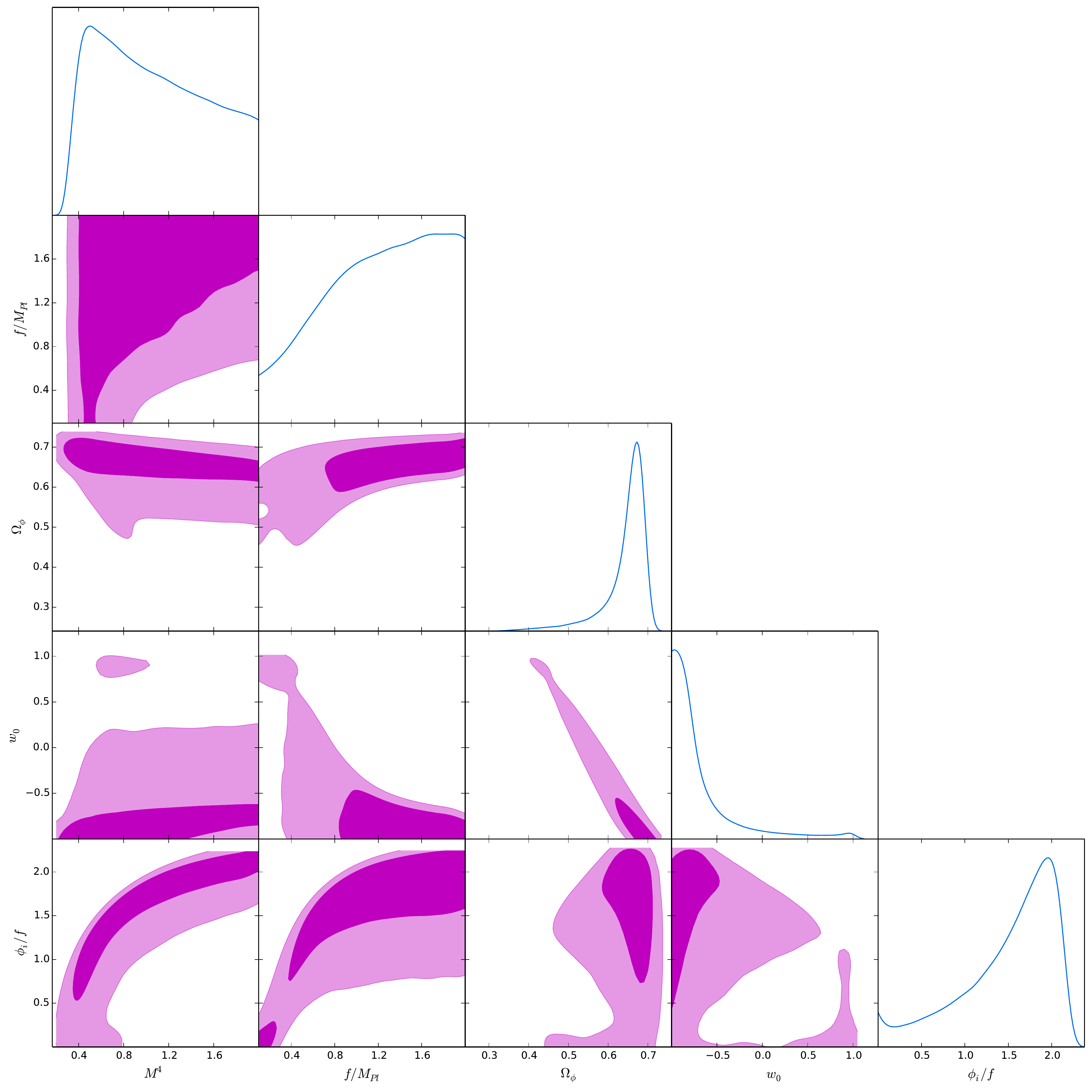}
       \caption{The 2D contours of the {\it Planck}-only constraints for the PNGB model with potential Eq.~(\ref{potential}).}  \label{fig:qLargePlanck}
\end{figure}

The results of the full-mission {\it Planck} observations of temperature and polarization aniso\-tropies of the CMB radiation are widely considered to be the most reliable dataset for constraining cosmological models. Moreover, for the standard six-parameter cosmological model they are sufficient on their own to fix all the parameters accurately. Here we test whether this latter statement remains true with the more general dark energy model by comparing results using {\it Planck} 2015 only with those of the full dataset described earlier. In Fig.~\ref{fig:qLargePlanck} a triangular plot with the  68\% and 95\% confidence contours using the {\it Planck} dataset is shown. The parameters displayed are the same as those of Fig.~\ref{fig:qStandardAll}, but note that in some cases the axis ranges have had to be extended.
Typically the constraints obtained by this particular analysis are considerably weaker than those of Sec.~\ref{PNGB:results}. The most striking example of this is the width of the potential $f/M_{\rm Pl}$; the {\it Planck} data alone is unable to constrain this within the prior at a confidence range of 95\%. A somewhat better constraint is attained for the width of the potential $M^4$, where at 68\%, $M^4 = 1.1_{-0.7}^{+0.3}$. This result is linked to the field density parameter limit of $\Omega_{\phi} = 0.64_{-0.01}^{+0.05}$, because of these two quantities' relation in the convergence of the code as mentioned in Sec.~\ref{PNGB:cosmomc}. There is a noticeable drop in the mean value of $\Omega_{\phi}$ against the full dataset result of $\Omega_{\phi} = 0.69$; as models with a lower potential amplitude are accepted and the field is allowed to roll for a wider variety of initial conditions.

Most notably, the present value of the equation of state is not well determined, with a 95\% confidence upper limit of $w_0 < 0.27$. This result is not very useful for a quintessence scenario, given that it does not even require the condition of $w_0 < -1/3$ for a Universe in accelerated expansion at present. There is also an unexpected portion of model space around $w_0 = +1$, which is not compatible with broader datasets. We conclude that a CMB-only analysis of the PNGB model is insufficient to distinguish between the permitted parameter combinations, and hence it is necessary to include the geometric data compiled at much lower redshift values. This was found also in Ref.~\cite{PlanckDE15} for the ($w_0$,$w_a$) parametrization of dark energy.
\subsection{ \label{PNGB:discussion:priors} Choice of inverse prior for the PNGB potential width}
In light of the data's inability to constrain high values of $f$, we assess the prior dependence of our results. We first carried out an analysis which simply extended the upper limit of the uniform prior on $f$ to 4 instead of 2, which just has the effect of admitting extra models at the higher values of $f$ which make predictions very similar to the cosmological constant. As the prior contains more models of this type, constraints on models away from this limit tighten somewhat, which already indicates a significant prior dependence to any constraints which are quoted.

More importantly, however, there is no particular motivation for choosing a prior uniform in $f$, which did not allow for a clear discrimination between models with a narrow width of $f/M_{\rm{Pl}} < 1$. In order to explore this we changed the parameter with uniform prior from $f/M_{\rm{Pl}}$ to $1/(f/M_{\rm{Pl}})$. We take the prior on this to be ($10^{-4}$, $5$), leaving the rest of our parameters with the same meanings as in Sec.~\ref{PNGB:results}. This then means that the cosmological constant limit occupies a finite region of parameter space near $1/f \simeq 0$, rather than the potentially-infinite region when placing a uniform prior on $f$.
\begin{figure}[t]
        \centering
        \includegraphics[width=0.99 \textwidth  ]{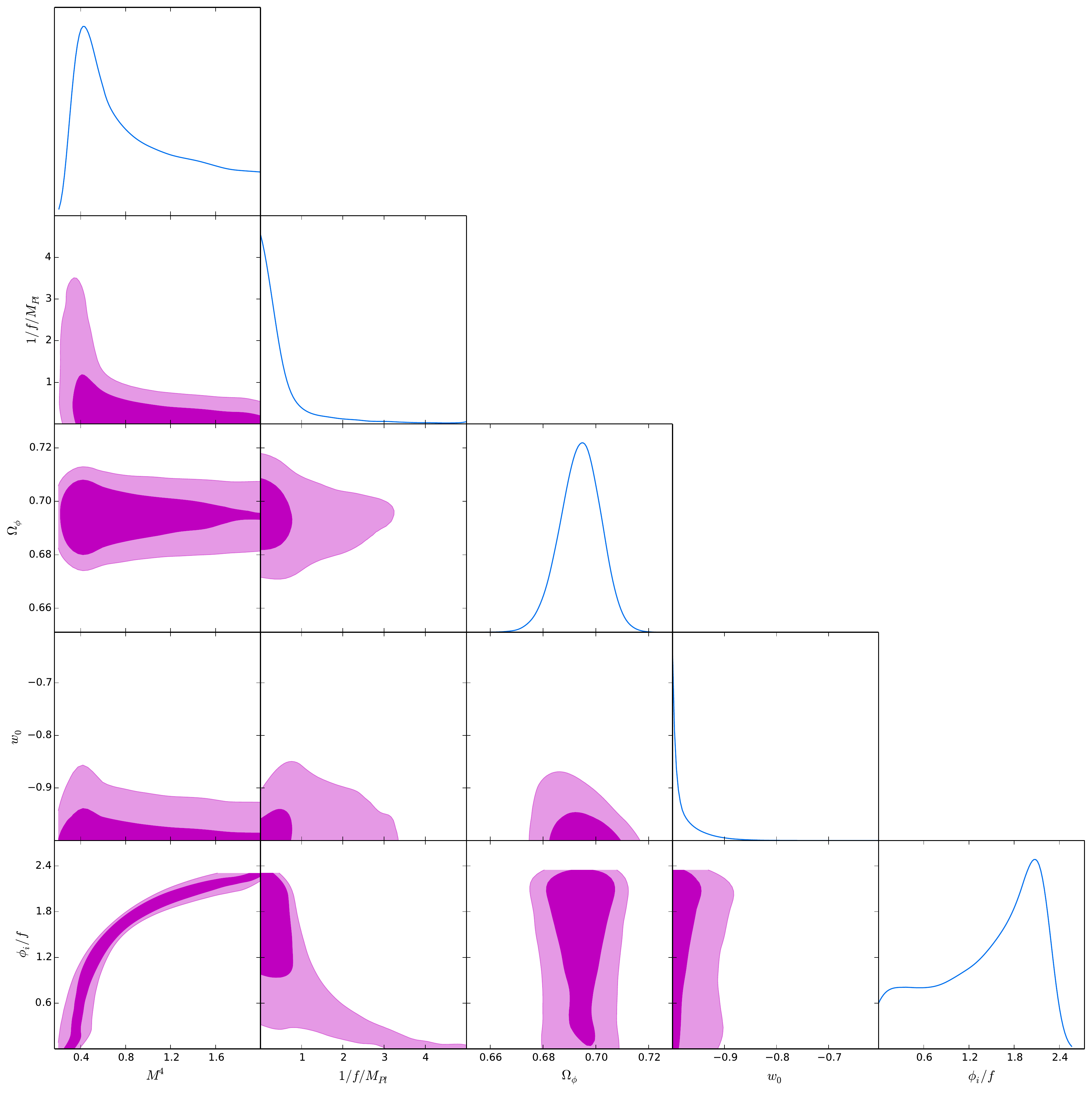}               
       \caption{The 2D contours of the combined JLA + BAO + HST + {\it Planck} 2015 constraints for the PNGB model with potential Eq.~(\ref{potential}) and parametrization uniform in $M_{\rm{Pl}}/f$ instead of $f/M_{\rm{Pl}}$. The rest of the parameter definitions are the same as in Fig.~\ref{fig:qStandardAll}.}  \label{fig:qOverfPlanck}
\end{figure}
Figure~\ref{fig:qOverfPlanck} shows the results of this new choice of prior, where $1/(f/M_{\rm{Pl}}) = 2.3$ is the upper limit found at a 95\% confidence level. This corresponds to a limit on $f$ which is not very different to the one we found in Sec.~\ref{PNGB:results}, though of course the changed prior modifies the overall shape of the posterior. The lower prior limit of $10^{-4}$ is close to zero, and therefore indistinguishable by the data due to its closeness to the cosmological constant case. 

The rest of the probability distributions are in good agreement as those of Sec.~\ref{PNGB:results}, indicating that they do not have much sensitivity to the choice of prior. The present equation of state $w_0$ is a good example of this insensitivity, its limit changing only slightly. However, the potential amplitude $M^4$ does become less constrained at its upper values. This is because our alternative prior places less prior weight on models which are very cosmological-constant-like, hence permitting a somewhat wider range of $M^4$ than our original prior choice.
\subsection{ \label{PNGB:discussion:previous} Comparison with previous results}
In this section we compare our results to those of previous authors. Similar studies to Fig.~\ref{fig:qStandardAll} can be found in Refs.~\cite{Abrahamse, Ng, Kawasaki, Dutta} and comparison is straightforward after noting a recurrent difference in the definition of the variable $M^4$ between those works and ours. As a typical example, Fig.~2 in Ref.~\cite{Dutta} plots
\begin{equation}
\left[\frac{f}{10^{18} \, {\rm GeV}}, \frac{\mu} {\sqrt{h/0.65}\times 10^{-3} \, {\rm eV}}\right],
\label{fmiu}
\end{equation}
where $\mu^4$ is the amplitude of the potential, $h$ is the Hubble parameter and $f$ is the width of the potential, same as in our case. This choice arises from the fact that the value of the critical density is $\rho_{\mathrm{crit}} = (3 \sqrt{h}$  $10^{-3} {\rm eV})^4$. The fixed value of $h = 0.65$ specified by the authors gives $\rho_{\mathrm{crit}} = (2.5 \times 10^{-3} {\rm eV})^4$.  To translate Eq.~(\ref{fmiu}) into our own definition of $M^4$ requires
\begin{equation}
\mu^4 = \rho_\mathrm{crit} M^4 
\label{miu}
\end{equation}
where $\mu$ is the amplitude of the potential in previous references. This choice makes their $\mu / (\sqrt{h}$ $10 ^{-3} \, {\rm eV} ) \approx 2 $, which is the minimum amplitude of the potential that would give the observed dark energy density. Therefore, $M^4_{\rm min} \approx 0.34$ in our Fig.~\ref{fig:qStandardAll} is equivalent to $\mu / (\sqrt{h}$ $10^{-3} \,{\rm eV}) \approx 2.29$ in Fig.~2 of Ref.~\cite{Dutta}. This somewhat larger $M_{\rm{min}}^4$ is expected, as Ref.~\cite{Dutta} imposes a hard limit $\Omega_{\phi} \geq 0.7$ on its models, whereas our data combination constraints prefer a smaller $\Omega_{\phi}$.  A very similar description can be made about the results of the rest of the aforementioned references. 

Our analysis results in a significant improvement on the constraint on $\Omega_{\rm \phi}$ in relation to Ref.~\cite{Dutta}; however, the restriction in the same work to $w_0 < -0.965$, citing the inability of the data to provide a better constraint on this parameter, is invalidated as a significantly larger range for the present value of the equation of state is shown in Fig.~\ref{fig:qStandardAll}. Concerning the slope of the scalar field, the allowed amount of rolling in $\phi_\mathrm{i}/f$ is in very good agreement with this reference. 

Regarding our results for the equation of state parameterization scheme, a comparison can be made with Fig.~7 in Ref.~\cite{Chiba}. These authors also explore models with $w_0 < -1$; however, because quintessence theories do not extend to the phantom domain, we excluded that possibility in the prior of our parameter space from the start. Another noticeable difference is their use of two datasets and their respective confidence levels; one using BOSS data and one without it. When adding BOSS they obtain a confidence contour that is entirely in the region $w_0 < -1$, but when it is omitted, a small area with $w_0 > -1$ is still allowed.  

The shape of the confidence region in Ref.~\cite{Chiba} differs greatly from the result displayed in our Fig.~\ref{fig: K_w_2D}. This is because that article adopted uniform priors in the equation of state parameters $w_0$ and $K$. Here, instead, these parameters are derived from a sampling based on assumption of the underlying PNGB potential. This induces a prior on those parameters, and particularly on $K$, which is very far from uniform, i.e.~the PNGB model realises a very different model ensemble from that assumed in Ref.~\cite{Chiba}. A simple consequence of using an underlying potential is that $w_0 < -1$ is not permitted, but the induced prior on $K$ also leads to a substantially different allowed region. Hence caution is required in using the equation of state approach to assess the viability of explicit thawing quintessence models such as PGNB.
\section{ \label{PNGB:conclusions} Conclusions}
In this paper we obtained a probability distribution for  PNGB quintessence model parameters, which corresponds to the thawing type. Our analysis was carried out using the codes CosmoMC/CAMB, version July 2015. The constraints on the amplitude $M^4$, initial slope of the field $\phi_\mathrm{i}/f$, and width $f/M_{\rm{Pl}}$ of the potential Eq.~(\ref{potential}) show good agreement with earlier analyses, with some reduction in the permitted range of values as compared to them. Our results show the continued viability of the $\Lambda$CDM scheme, while showing the extent to which models with a present equation of state value larger than $w_0 = -1$ and a field evolving away from the top of the potential (\ref{potential}), corresponding to thawing quintessence, remain acceptable.

We then studied the approximate equation of state (\ref{ThawingEqState}), which has been applied in Refs.~\cite{Chiba,DuttaAndScherrer}. Under the PNGB assumption, we find much tighter constraints on the curvature parameter $K$ than are found when uniform priors are adopted on the equation of state parameters. This difference highlights a strong ongoing prior dependence from the way thawing dark energy is modelled, which current data are not strong enough to override. We explored the parameter space with different dataset combinations to test their effectiveness. We discussed the advantages of a different prior choice and compared with the standard parametrization.

Overall we conclude that the current data are indeed able to meaningfully constrain the PNGB model, and in particular force its behaviour to be close to the cosmological constant limit. However, in detail there remains a significant prior dependence on the constraints obtained, as highlighted by switching the prior from being uniform on $f$ to being uniform on $1/f$. In absence of any clear theoretical guidance on the appropriate form of prior, this dependence needs to be kept in mind in interpreting constraints. Hopefully future data will emerge with sufficient strength to overcome this prior dependence.

\section*{Acknowledgments}

V.S.-B.\ acknowledges funding provided by CONACyT and the University of Edinburgh, while  A.R.L.\ was supported by STFC under grant numbers ST/K006606/1 and ST/L000644/1. This work was undertaken on the COSMOS Shared Memory system at DAMTP, University of Cambridge operated on behalf of the STFC DiRAC HPC Facility. This equipment is funded by BIS National E-infrastructure capital grant ST/J005673/1 and STFC grants ST/H008586/1 and ST/K00333X/1. The authors thank Nelson Lima and Luis Ure\~na-L\'opez  for useful advice on details of the computational analysis.

\bibliographystyle{JHEP}
\bibliography{PNGB}

\providecommand{\href}[2]{#2}\begingroup\raggedright\begin{thebibliography}{10}

\bibitem{Riess}
A.~G. {Riess}, A.~V. {Filippenko}, P.~{Challis}, A.~{Clocchiatti},
  A.~{Diercks}, P.~M. {Garnavich} et~al., \emph{{Observational Evidence from
  Supernovae for an Accelerating Universe and a Cosmological Constant}},
  \href{http://dx.doi.org/10.1086/300499}{\emph{\aj} {\bf 116} (Sept., 1998)
  1009--1038}, [\href{http://arxiv.org/abs/astro-ph/9805201}{{\tt
  astro-ph/9805201}}].

\bibitem{Perlmutter}
S.~{Perlmutter}, G.~{Aldering}, G.~{Goldhaber}, R.~A. {Knop}, P.~{Nugent},
  P.~G. {Castro} et~al., \emph{{Measurements of {$\Omega$} and {$\Lambda$} from
  42 High-Redshift Supernovae}},
  \href{http://dx.doi.org/10.1086/307221}{\emph{\apj} {\bf 517} (June, 1999)
  565--586}, [\href{http://arxiv.org/abs/astro-ph/9812133}{{\tt
  astro-ph/9812133}}].

\bibitem{PlanckCos15}
{\scshape Planck} collaboration, P.~A.~R. Ade et~al., \emph{{Planck 2015
  results. XIII. Cosmological parameters}},
  \href{http://dx.doi.org/10.1051/0004-6361/201525830}{\emph{Astron.
  Astrophys.} {\bf 594} (2016) A13},
  [\href{http://arxiv.org/abs/1502.01589}{{\tt 1502.01589}}].

\bibitem{PlanckDE15}
{Planck Collaboration}, P.~A.~R. {Ade}, N.~{Aghanim}, M.~{Arnaud},
  M.~{Ashdown}, J.~{Aumont} et~al., \emph{{Planck 2015 results. XIV. Dark
  energy and modified gravity}},
  \href{http://dx.doi.org/10.1051/0004-6361/201525814}{\emph{Astron.
  Astrophys.} {\bf 594} (Sept., 2016) A14},
  [\href{http://arxiv.org/abs/1502.01590}{{\tt 1502.01590}}].

\bibitem{Copeland:2006wr}
E.~J. {Copeland}, M.~{Sami} and S.~{Tsujikawa}, \emph{{Dynamics of Dark
  Energy}},
  \href{http://dx.doi.org/10.1142/S021827180600942X}{\emph{International
  Journal of Modern Physics D} {\bf 15} (2006) 1753--1935},
  [\href{http://arxiv.org/abs/hep-th/0603057}{{\tt hep-th/0603057}}].

\bibitem{Tsuji}
S.~Tsujikawa, \emph{{Quintessence: A Review}},
  \href{http://dx.doi.org/10.1088/0264-9381/30/21/214003}{\emph{Class. Quant.
  Grav.} {\bf 30} (2013) 214003}, [\href{http://arxiv.org/abs/1304.1961}{{\tt
  1304.1961}}].

\bibitem{Stebbins}
J.~A. Frieman, C.~T. Hill, A.~Stebbins and I.~Waga, \emph{{Cosmology with
  ultralight pseudo Nambu-Goldstone bosons}},
  \href{http://dx.doi.org/10.1103/PhysRevLett.75.2077}{\emph{Phys. Rev. Lett.}
  {\bf 75} (1995) 2077--2080},
  [\href{http://arxiv.org/abs/astro-ph/9505060}{{\tt astro-ph/9505060}}].

\bibitem{Dutta}
K.~Dutta and L.~Sorbo, \emph{{Confronting pNGB quintessence with data}},
  \href{http://dx.doi.org/10.1103/PhysRevD.75.063514}{\emph{Phys. Rev.} {\bf
  D75} (2007) 063514}, [\href{http://arxiv.org/abs/astro-ph/0612457}{{\tt
  astro-ph/0612457}}].

\bibitem{Waga}
I.~Waga and J.~A. Frieman, \emph{{New constraints from high redshift supernovae
  and lensing statistics upon scalar field cosmologies}},
  \href{http://dx.doi.org/10.1103/PhysRevD.62.043521}{\emph{Phys. Rev.} {\bf
  D62} (2000) 043521}, [\href{http://arxiv.org/abs/astro-ph/0001354}{{\tt
  astro-ph/0001354}}].

\bibitem{Ng}
S.~C.~C. Ng and D.~L. Wiltshire, \emph{{Properties of cosmologies with
  dynamical pseudo Nambu-Goldstone bosons}},
  \href{http://dx.doi.org/10.1103/PhysRevD.63.023503}{\emph{Phys. Rev.} {\bf
  D63} (2001) 023503}, [\href{http://arxiv.org/abs/astro-ph/0004138}{{\tt
  astro-ph/0004138}}].

\bibitem{Kawasaki}
M.~Kawasaki, T.~Moroi and T.~Takahashi, \emph{{Cosmic microwave background
  anisotropy with cosine type quintessence}},
  \href{http://dx.doi.org/10.1103/PhysRevD.64.083009}{\emph{Phys. Rev.} {\bf
  D64} (2001) 083009}, [\href{http://arxiv.org/abs/astro-ph/0105161}{{\tt
  astro-ph/0105161}}].

\bibitem{Abrahamse}
A.~Abrahamse, A.~Albrecht, M.~Barnard and B.~Bozek, \emph{{Exploring parameter
  constraints on quintessential dark energy: The pseudo-Nambu-Goldstone-boson
  model}}, \href{http://dx.doi.org/10.1103/PhysRevD.77.103503}{\emph{Phys.
  Rev.} {\bf D77} (2008) 103503}, [\href{http://arxiv.org/abs/0712.2879}{{\tt
  0712.2879}}].

\bibitem{DuttaAndScherrer}
S.~Dutta and R.~J. Scherrer, \emph{{Hilltop Quintessence}},
  \href{http://dx.doi.org/10.1103/PhysRevD.78.123525}{\emph{Phys. Rev.} {\bf
  D78} (2008) 123525}, [\href{http://arxiv.org/abs/0809.4441}{{\tt
  0809.4441}}].

\bibitem{ChibaFirst}
T.~Chiba, \emph{{Slow-Roll Thawing Quintessence}},
  \href{http://dx.doi.org/10.1103/PhysRevD.80.109902,
  10.1103/PhysRevD.79.083517}{\emph{Phys. Rev.} {\bf D79} (2009) 083517},
  [\href{http://arxiv.org/abs/0902.4037}{{\tt 0902.4037}}].

\bibitem{Chiba}
T.~Chiba, A.~De~Felice and S.~Tsujikawa, \emph{{Observational constraints on
  quintessence: thawing, tracker, and scaling models}},
  \href{http://dx.doi.org/10.1103/PhysRevD.87.083505}{\emph{Phys. Rev.} {\bf
  D87} (2013) 083505}, [\href{http://arxiv.org/abs/1210.3859}{{\tt
  1210.3859}}].

\bibitem{Hlozek:2014lca}
R.~Hlozek, D.~Grin, D.~J.~E. Marsh and P.~G. Ferreira, \emph{{A search for
  ultralight axions using precision cosmological data}},
  \href{http://dx.doi.org/10.1103/PhysRevD.91.103512}{\emph{Phys. Rev.} {\bf
  D91} (2015) 103512}, [\href{http://arxiv.org/abs/1410.2896}{{\tt
  1410.2896}}].

\bibitem{CAMB}
A.~Lewis, A.~Challinor and A.~Lasenby, \emph{{Efficient computation of CMB
  anisotropies in closed FRW models}},
  \href{http://dx.doi.org/10.1086/309179}{\emph{Astrophys. J.} {\bf 538} (2000)
  473--476}, [\href{http://arxiv.org/abs/astro-ph/9911177}{{\tt
  astro-ph/9911177}}].

\bibitem{PlanckLikelihood}
{\scshape Planck} collaboration, N.~Aghanim et~al., \emph{{Planck 2015 results.
  XI. CMB power spectra, likelihoods, and robustness of parameters}},
  \href{http://dx.doi.org/10.1051/0004-6361/201526926}{\emph{Astron.
  Astrophys.} {\bf 594} (2016) A11},
  [\href{http://arxiv.org/abs/1507.02704}{{\tt 1507.02704}}].

\bibitem{CosmoMC}
A.~Lewis and S.~Bridle, \emph{{Cosmological parameters from CMB and other data:
  A Monte Carlo approach}},
  \href{http://dx.doi.org/10.1103/PhysRevD.66.103511}{\emph{Phys. Rev.} {\bf
  D66} (2002) 103511}, [\href{http://arxiv.org/abs/astro-ph/0205436}{{\tt
  astro-ph/0205436}}].

\bibitem{Banks}
T.~Banks, M.~Dine, P.~J. Fox and E.~Gorbatov, \emph{{On the possibility of
  large axion decay constants}},
  \href{http://dx.doi.org/10.1088/1475-7516/2003/06/001}{\emph{JCAP} {\bf 0306}
  (2003) 001}, [\href{http://arxiv.org/abs/hep-th/0303252}{{\tt
  hep-th/0303252}}].

\bibitem{JLA}
{\scshape SDSS} collaboration, M.~Betoule et~al., \emph{{Improved cosmological
  constraints from a joint analysis of the SDSS-II and SNLS supernova
  samples}}, \href{http://dx.doi.org/10.1051/0004-6361/201423413}{\emph{Astron.
  Astrophys.} {\bf 568} (2014) A22},
  [\href{http://arxiv.org/abs/1401.4064}{{\tt 1401.4064}}].

\bibitem{HST}
A.~G. Riess, L.~Macri, S.~Casertano, H.~Lampeitl, H.~C. Ferguson, A.~V.
  Filippenko et~al., \emph{{A 3\% Solution: Determination of the Hubble
  Constant with the Hubble Space Telescope and Wide Field Camera 3}},
  \href{http://dx.doi.org/10.1088/0004-637X/732/2/129,
  10.1088/0004-637X/730/2/119}{\emph{Astrophys. J.} {\bf 730} (2011) 119},
  [\href{http://arxiv.org/abs/1103.2976}{{\tt 1103.2976}}].

\bibitem{BAO11}
{\scshape BOSS} collaboration, L.~Anderson et~al., \emph{{The clustering of
  galaxies in the SDSS-III Baryon Oscillation Spectroscopic Survey: baryon
  acoustic oscillations in the Data Releases 10 and 11 Galaxy samples}},
  \href{http://dx.doi.org/10.1093/mnras/stu523}{\emph{Mon. Not. Roy. Astron.
  Soc.} {\bf 441} (2014) 24--62}, [\href{http://arxiv.org/abs/1312.4877}{{\tt
  1312.4877}}].

\end{thebibliography}\endgroup

\end{document}